\documentclass[runningheads]{llncs}
\usepackage[T1]{fontenc}
\usepackage{graphicx}
\usepackage{listings}
\usepackage{makecell}
\usepackage{todonotes}
\usepackage{multirow}
\newcommand{\listingsfont}{\ttfamily}
\lstdefinestyle{pseudocode}{                                                                                                                    basicstyle=\small\ttfamily,
    mathescape=true,
    columns=fullflexible,
    keepspaces=true,
    xleftmargin=1em,
    frame=single,
  }   

\begin{document}
\title{Exploring spectral element methods on the Tenstorrent RISC-V accelerator}
%
%
\author{Daniyal Arshad\orcidID{0009-0008-6523-7582} \and
Nick Brown\orcidID{0000-0003-2925-7275}}
\authorrunning{D. Arshad et al.}
%
\institute{EPCC, The University of Edinburgh, 47 Potterrow, Edinburgh, UK}
\maketitle              
\begin{abstract}
The growing availability of commodity RISC-V hardware has sparked interest in its use for High Performance Computing (HPC), with PCIe accelerator cards offering a practical near-term pathway to adoption. The Tenstorrent Wormhole is one example, with dedicated vector and matrix units across 128 Tensix cores, and is widely available. In this paper, we explore porting the AX kernel of Nekbone, a widely used HPC mini-application derived from the Gordon Bell Prize-winning Nek5000 spectral element solver, onto the Wormhole accelerator. This kernel evaluates the Poisson operator, and we describe the mapping of the algorithm onto the Tensix. The initial performance results reveal that the host-side data transposition, required for the z-direction gradient computation, is a severe bottleneck. Consequently, we investigated two optimisation strategies that yield dramatic improvements, achieving 242.97 GFLOPS for 100000 elements across 128 Tensix cores, outperforming a 24-core Xeon Platinum CPU and drawing approximately 7 times less power. 

\keywords{RISC-V \and Tenstorrent \and Nekbone \and Spectral element method}
\end{abstract}
\section{Introduction}

The recent availability of high-core-count commodity RISC-V CPUs \cite{brown2025risc} has generated growing interest in the potential of RISC-V architectures for High Performance Computing (HPC) \cite{brown2023risc}. Nevertheless, the broader ecosystem has yet to mature sufficiently to support fully CPU-based RISC-V supercomputers. In the near term, a more incremental pathway to adoption is therefore expected through the use of RISC-V-based PCIe accelerator cards. These devices offer the key advantage of seamless integration into existing x86 or AArch64 systems as modular add-ons. Several vendors are actively developing such accelerator cards, with many targeting artificial intelligence (AI) and machine learning (ML) workloads in response to the ongoing AI surge. Notably, each vendor has adopted a distinct design philosophy shaped by its own priorities, underscoring the inherent flexibility of the RISC-V standard, which enables hardware designers to tailor implementations to their specific requirements.

Irrespective of whether they were originally conceived for ML or HPC applications, RISC-V accelerator hardware fundamentally supplies the foundational components required to accelerate mathematical operations. As a result, these technologies present a compelling opportunity for the HPC community. One prominent example is the Wormhole accelerator card developed by Tenstorrent. As of 2026, this commercially available card is still among the few publicly available RISC-V-based accelerators and can be purchased at a modest price point. Its affordability and availability render it suitable not only for integration into leading supercomputers but also for smaller-scale HPC systems and even high-performance workstations. In addition, Tenstorrent has open-sourced their entire software stack and actively collaborates with the wider research and development community.

In this paper, we explore porting the AX kernel of Nekbone, a popular HPC mini-app, which captures the basic structure of the Nek5000 spectral element method based application onto a Tenstorrent Wormhole RISC-V accelerator. AX evaluates the Poisson operator, accounts for approximately 75\% of the code's runtime, and suffers from memory bottlenecks on the CPU. As it is heavily matrix multiplication based, our hypothesis is that not only will it map well to the Tensix's matrix unit, but furthermore the memory overhead could be significantly reduced due to the architecture's design. This paper is structured as follows; Section \ref{sec:background} describes the Tenstorrent Tensix architecture and Nekbone mini-app in more detail before we explore the steps undertaken to port the code in Section \ref{sec:porting}. Section \ref{sec:initial} reports initial performance results and based upon the observation that one part of the ported code is a significant bottleneck, we then explore optimisation strategies in Section \ref{sec:optimise} reporting performance against a CPU, GPU and FPGA. Lastly, Section \ref{sec:conc} draws conclusions and discusses future work.

\section{Background}
\label{sec:background}

\begin{figure}[htb]
\centering
 \includegraphics[width=\columnwidth]{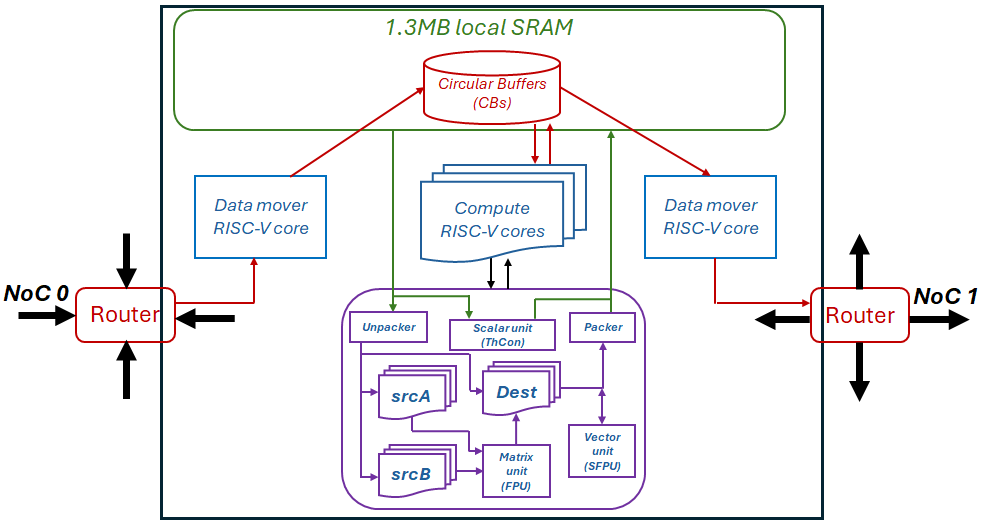}
\caption{A single Tensix core in the Wormhole accelerator, containing five RISC-V \emph{baby cores}, 1.3MB of local SRAM memory, a compute engine and two routers to the Network on Chip (NoC). From \cite{brown2025exploring}}	
\label{fig:tensix}
\end{figure}

The Tensix architecture is illustrated in Figure \ref{fig:tensix} where each Tensix core comprises five RISC-V processors, referred to as baby cores, local SRAM, two routers interfacing with independent Networks-on-Chip (NoCs), and a dedicated compute co-processor. Of the five baby cores, one is responsible for ingressing data into the Tensix core, one for egressing data, and the remaining three drive the co-processor. The co-processor provides scalar (ThCon), vector (SFPU), and matrix (FPU) units supporting a range of numerical precisions up to FP32, but with the FPU only providing Bfloat-16 (BF16) and TensorFloat-32 (TF32). The \emph{srcA} and \emph{srcB} registers are each 4 KiB in capacity, accommodating up to 2048 BF16 values, while the \emph{dst} register is 32 KiB in size and partitioned into 16 segments \cite{corsix}. To prevent the RISC-V compute baby cores from becoming a throughput bottleneck, data values are never routed through them directly, and instead the co-processor is instructed to access SRAM directly.

In this work, we use the matrix unit exclusively and this supports only a limited set of operations: matrix multiplication, transposition, reduction, and element-wise arithmetic (addition, subtraction, and multiplication). The vector unit, by contrast, is considerably more versatile and beyond standard binary arithmetic, also supports operations including squares, square roots, cosine, tangent, sine, and conditionals. Programmers may also write custom code directly for the SFPU, which the compiler will then vectorise. The matrix unit can perform up to a theoretical limit of 4.096 TFLOPS \cite{tt-isa-doc}, and in this work we use the matrix multiplication and transposition functionality of the FPU.

The \emph{Circular Buffer} (CB) abstraction represents the data held in the local SRAM. CBs are First-In First-Out (FIFO) queues operating on a producer-consumer model. Following this approach the producer requests a page within the CB, populates it, and pushes it out for consumption. The consumer blocks until a page becomes available, and when it does this is then processed and freed with the memory being able to be reused by the producer. By combining memory management with synchronisation, CBs facilitate coordination between the RISC-V baby cores. An example is a data mover core requesting a CB page, loading it with data fetched from external DRAM, and pushing it, after which the RISC-V compute cores consume the CB, with the \emph{UNPACK} core directing the \emph{unpacker} in the co-processor to copy the page's contents into a target register.

This work targets a Wormhole n300, which houses 128 Tensix cores distributed across two 64-core chips clocked at 1 GHz. The board provides a total of 24 GB of GDDR6 external memory, organised into twenty-four 1 GB banks, with twelve banks attached directly to each chip \cite{n300-datasheet}. TT-Metalium (\emph{tt-metal}) is Tenstorrent's low-level programming SDK, offering direct hardware access through an API designed for kernel development. Primitives for data movement, compute engine control, and CB management are provided and developers write three kernels: one for each of the two data movement RISC-V cores and one for the compute cores.

Previous work \cite{brown2024accelerating} explored stencil applications on Tenstorrent's first generation, the Grayskull e150, and both FFTs \cite{brown2025exploring} and gravitational N-Body Simulation \cite{almerol2025accelerating} on the Wormhole. Throughout this existing work, it has been demonstrated that performance is generally similar, and sometimes better, to server grade CPUs but with significantly reduced power draw. However, the major challenge is tuning applications for the Tenstorrent architecture where, for example, in \cite{brown2024accelerating} the final version of the code was around 160 times faster on the Grayskull than the initial, naive, kernel implementation.

\subsection{Nekbone}
\label{sec:nekbone}
Nekbone \cite{ivanov2015evaluation} is a mini-application derived from Nek5000 \cite{nek5000-web-page}, a high-order incompressible Navier–Stokes solver that employs the spectral element method. Nek5000, a recipient of the Gordon Bell Prize, along with its Nekbone proxy application, has achieved widespread adoption within the HPC community. Nekbone solves a standard Poisson equation via the Conjugate Gradient (CG) iterative method with a simple preconditioner on either a block or linear geometry. As this constitutes the principal computational kernel of Nek5000, Nekbone serves as a valuable vehicle for investigating the core algorithmic characteristics shared by Nek5000 and numerous other HPC applications. Consequently, insights gained from accelerating Nekbone are directly transferable to Nek5000 and are broadly applicable to the wider class of HPC codes that employ analogous computational strategies.

The Nekbone solution phase comprises CG iterations, each involving vector operations, matrix–matrix multiplications, nearest-neighbour communication, and MPI Allreduce operations. Among these, the dominant computational bottleneck is the AX kernel, which evaluates the Poisson operator and accounts for approximately 75\% of total runtime. Accordingly, this work focuses exclusively on the AX kernel, in which all computations are performed on an element-by-element basis, with each element defined by a specific polynomial order configuration. The code is configured with a number of elements in three dimensions, and a polynomial order (NX) whose cube determines the number of grid points per element. The AX kernel represents a particularly demanding computational pattern where each element requires a series of relatively small BLAS operations, and the cumulative effect of executing a large number of such small operations is typically less efficient than performing a single, equivalently sized BLAS operation \cite{dongarra2017design}.

Listing \ref{lst:ax} sketches pseudocode of the AX kernel, which is itself written in Fortran 77. It accepts \emph{nx} which is the polynomial order, \emph{D} which is the spectral differentiation matrix of size nx by nx, \emph{G} which is the geometric factor matrix and we assume is the identity matrix in this work, and \emph{u} which is an input field. Gradients for the \emph{x}, \emph{y} and \emph{z} dimensions are first calculated and stored as intermediate results \emph{ur}, \emph{us} and \emph{ut} respectively. Geometric factors are then applied and an accumulation is undertaken for each z layer, with the resulting \emph{w} field returned to the caller. Throughout this work we set the polynomial order, NX, to be 16 which is the largest commonly adopted configuration size and involves 96 matrix multiplications per element, with each element being independent of each other.

  \begin{lstlisting}[style=pseudocode, caption={AX operator for a single spectral element.}, label={lst:ax}]                                                                                      
  procedure AX(nx, D, G, u)
      // u is a field on an nx $\times$ nx $\times$ nx element grid
      // D is the 1D spectral differentiation matrix (nx $\times$ nx)
      // G is the 3 $\times$ 3 symmetric geometric factor matrix

      // Directional gradients
      for each z-layer:
          $u_r$(z) $\leftarrow$ D $\cdot$ u(z)          // x-gradient: matrix $\times$ layer
          $u_s$(z) $\leftarrow$ u(z) $\cdot$ D$^\mathsf{T}$         // y-gradient: layer $\times$ matrix

      for each (x,y) point:
          $u_t$(x,y,:) $\leftarrow$ D $\cdot$ u(x,y,:)  // z-gradient: matrix $\times$ column

      // Apply geometric factors
      for each point (x,y,z):
          [$u_r$, $u_s$, $u_t$] $\leftarrow$ G $\cdot$ [$u_r$, $u_s$, $u_t$]

      // Accumulate with transposed operator
      for each z-layer:
          w(z) $\leftarrow$ D$^\mathsf{T}$ $\cdot$ $u_r$(z) + $u_s$(z) $\cdot$ D + $u_t$(z) $\cdot$ D
      return w
  \end{lstlisting}   

The polynomial order (NX) of 16 requires 831,488 floating-point operations per element. \footnote{With $ G{=}I $ (geometric scaling omitted), the per-element cost is 794,624 FLOPS (used for both Tenstorrent and CPU FP32/FP64 in this work), reducing compute by 36864 FLOPS ($\approx$ 4.4\%).} Furthermore, different parts of the kernel are bound by different limits, some aspects of which are memory bound, and others compute bound. Previous work \cite{brown2020exploring} profiled this AX kernel on an Intel Xeon Platinum Cascade Lake (8260M) CPU and when running on one core, only 45\% of the cycles were undertaking useful work, with over 26\% stalling due to memory issues. When the code was scaled up to all 24 CPU cores, a speed up of less than 12 times was obtained, and on average a single core only completed useful work for 27\% of the cycles and over 56\% of cycles were stalled due to memory issues. Furthermore, when running on all 24 cores the system reported that memory bandwidth was at 100\% utilisation. This indicates that on the CPU this code is bound by memory accesses, stalling when running on a single core due to poor caching behaviour, and as one increases the number of cores the code also becomes memory bandwidth bound.

\section{Porting Nekbone's AX operator to the Tensix}
\label{sec:porting}
In this work, we focus on the Nekbone \emph{AX} operator, and a key challenge is how to map this most effectively to the Tensix architecture. The fundamental challenge is to map the algorithm to the details of the architecture, specifically calculations to the 32x32 fixed FPU tile size, and to also enable the calculations to run in a way that avoids data movement or reordering by the baby RISC-V cores on the device which has been shown to result in significant overhead \cite{brown2024accelerating}.

\begin{figure}[htb]
\centering
 \includegraphics[width=\columnwidth]{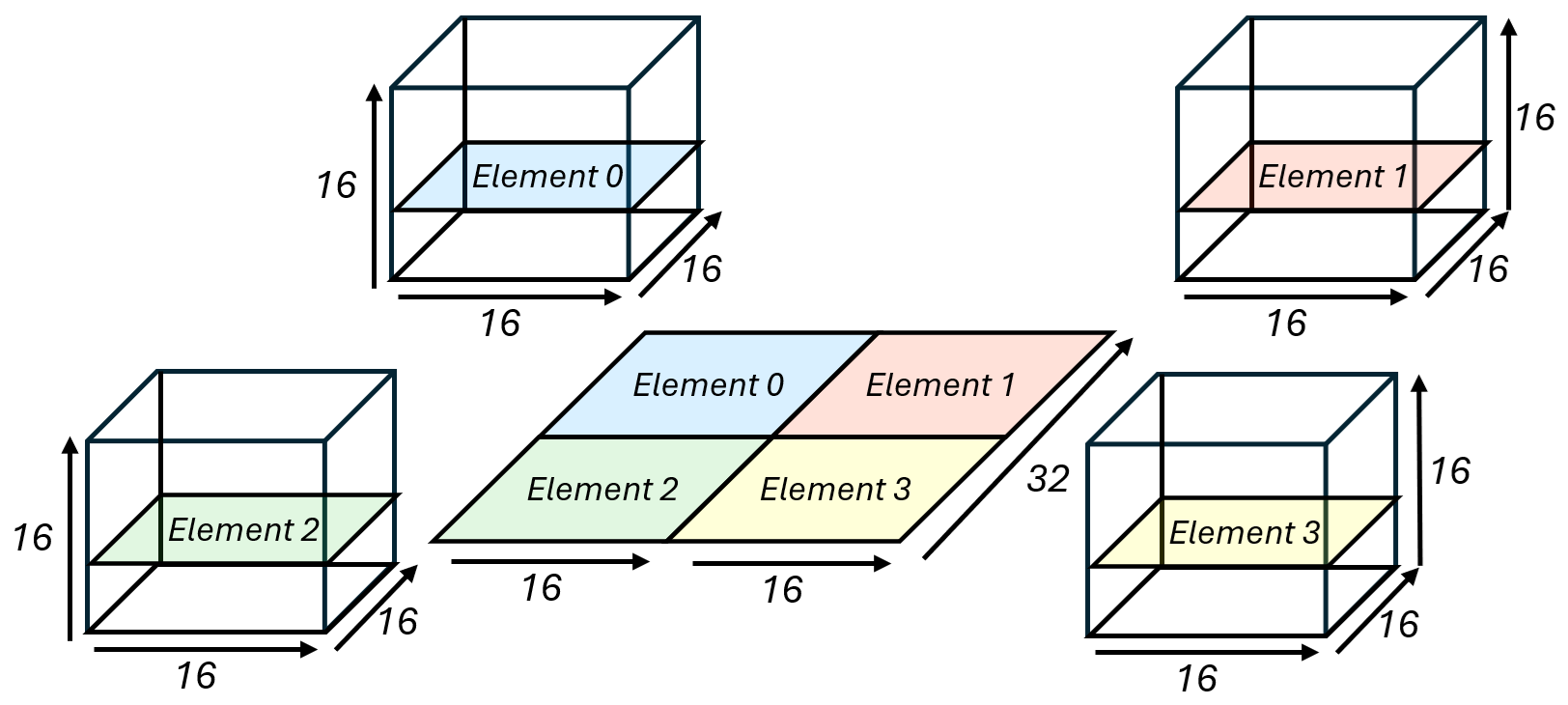}
\caption{Illustration of mapping 16x16 slice from four elements to each 32x32 FPU tile.}	
\label{fig:tile}
\end{figure}

As described in Section \ref{sec:nekbone}, each element is represented by three dimensions, the size of each dimension being that of the polynomial order, \emph{NX}. The total size of each field is therefore $NX^3$, for instance 4096 for NX=16, which is greater than the tile size of 32x32. Consequently, to avoid data reordering on the device, we perform calculations on a 2D slice by slice basis on the FPU, working up the column from one slice to the next. However, each slice is at most 16x16, as 16 is the largest polynomial order, whereas the FPU tile size is 32x32. Therefore, computing on one 2D slice at a time would waste significant compute resource. One solution would be to pack multiple slices for an element together, but that would result in data copying by the RISC-V baby cores on the device which must be avoided for performance. As illustrated in Figure \ref{fig:tile}, instead our solution is to pack multiple elements together in a single FPU tile. Figure \ref{fig:tile} provides an example where NX=16, where the x-y slices from four elements are packed into a single tile and the FPU then performs compute on each of these four elements concurrently, moving upwards through the columns working on slices from these four elements at each level. 

\begin{figure}[htb]
\centering
 \includegraphics[width=\columnwidth]{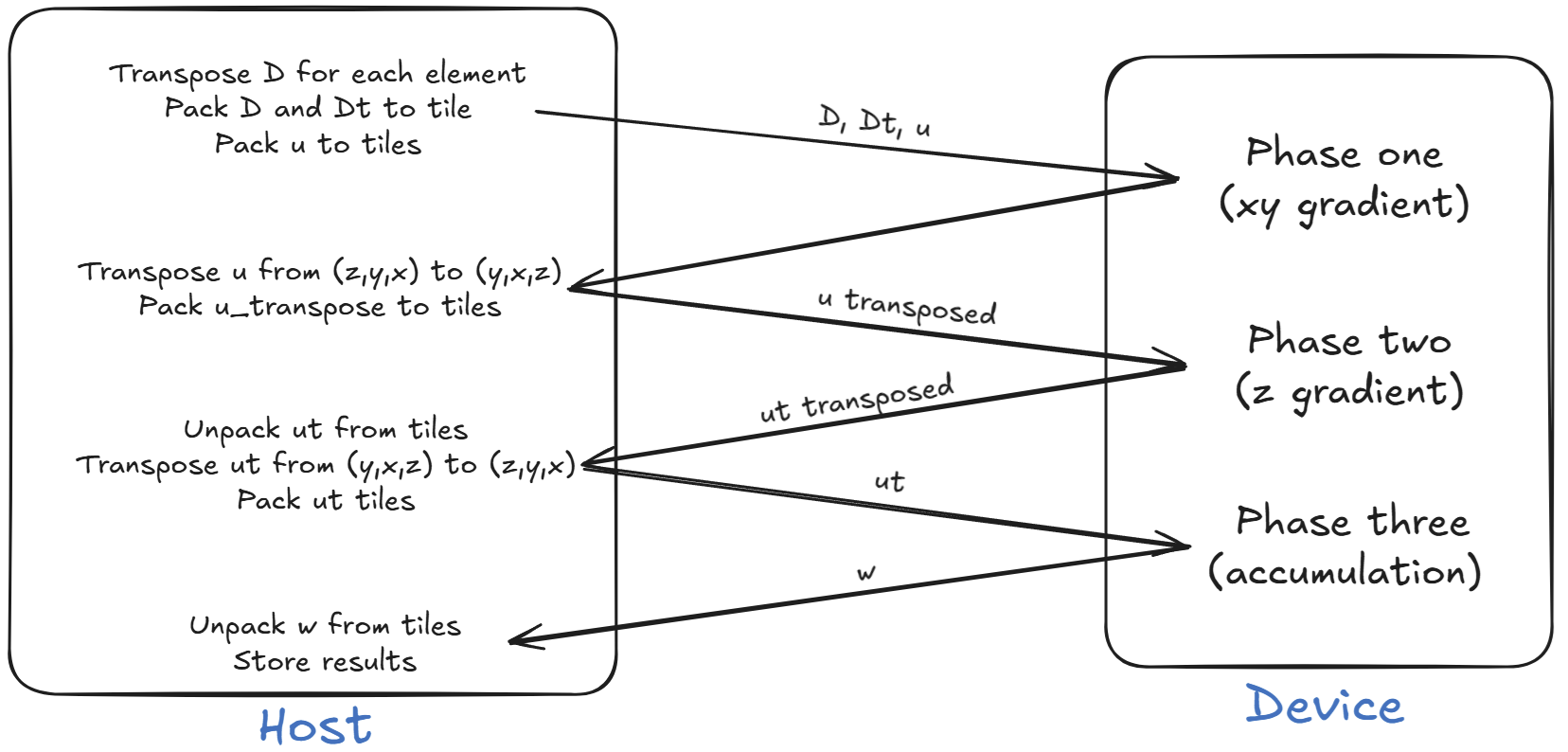}
\caption{Illustration of interaction between the host and device.}	
\label{fig:host-device}
\end{figure}

Figure \ref{fig:host-device} sketches how the code is structured, and it can be seen that there are three sequential phases launched onto the device, each corresponding to a part of the spectral element gradient computation. Each phase comprises three device-side kernels, one for data movement moving data from the DRAM into the Tensix core, one for compute, and one for data movement from the core back to DRAM. All data is represented as 32x32 BF16 tiles (2048 bytes each), which matches the hardware's native tile format.

The first phase computes the \emph{x} and \emph{y} direction gradients, which involves computing with the stiffness matrix operator. As sketched in Section \ref{sec:nekbone}, this is of the form \emph{$D^T$ G D}, where \emph{D} computes the directional derivatives to scatter field values into gradients, \emph{$D^T$} the adjoint operation to gather these weighted gradients back to the grid points, and in this work \emph{G} is the identity matrix. On the CPU, the code simply changes the index when accessing \emph{u} in the inner loop to achieve the transposition \emph{$D^T$}, avoiding any data copying. By contrast, on the Tenstorrent device compute must be performed via tile level matrix multiplication and-so there is not the same flexibility. Consequently, \emph{D} must be transposed on the host, and both \emph{D} and \emph{$D^T$} sent to the device along with the input \emph{u} field. 

\emph{D} and \emph{$D^T$} are of size \emph{NX} by \emph{NX} and values are constant across elements, therefore, the Tensix reading kernel loads the differentiation matrix \emph{D} and its transpose \emph{$D^T$} only once into Circular Buffers (CBs) and then reuses these across tiles. This same reading kernel streams the input field \emph{u} tile by tile, where each tile contains multiple elements (4 where NX=16) and the compute kernel instructs the FPU to perform two matrix multiplications, the first calculating \emph{ur = D × u} and the second \emph{us = u × $D^T$}. The results, \emph{ur} and \emph{us}, are then written to an output CB which the writer kernel consumes and stores in DRAM. These results, along with \emph{D} and \emph{$D^T$}, remain resident on the device as they will be used in phase three.

The second phase computes the \emph{z} direction gradient. However, whilst for the \emph{x} direction \emph{D} contracts along columns in a tile which involves matrix multiplication, and for the \emph{y} dimension \emph{D} contracts along rows which is matrix multiplication using \emph{$D^T$}, for the \emph{z} dimension \emph{D} needs to contract upwards across different tiles. The CPU code handles this by changing the loop order, however, on the Tenstorrent architecture this is not possible. Instead, as per Figure \ref{fig:host-device}, the host transposes the \emph{u} field from coordinates (z,y,x) to (y,x,z) before packing it to tiles and transferring these to the device which then executes the second phase. The device then performs a \emph{D × u\_transposed} multiplication across all tiles, with the results then written to DRAM. However, at this stage the results, \emph{ut}, follow (y,x,z) coordinates and are out of order for phase three which needs to consume them in (z,y,x). Consequently, \emph{ut} is transferred back to the host, individual elements from the tile are unpacked, converted from BF16 to FP64, transposed back to (z,y,x) orientation, converted back to BF16 and packed into tiles before sending this to the device for phase three.

The third phase performs the accumulation \emph{w = $D^T$ × ur + us × D + ut × D}, where \emph{D} and \emph{$D^T$} are again loaded into CBs only once by the reader kernel and result fields \emph{ur}, \emph{us} and \emph{ut} are streamed from DRAM to the compute kernel. The compute kernel then leverages the accumulation behaviour of the hardware's destination registers where three successive \emph{matmul\_tiles} calls targeting the same register produce the required sum without explicit addition operations. The output field, \emph{w} is stored by the compute kernel in an output CB which is then written to DRAM by the writer kernel. These results are then transferred back to the host where individual elements are unpacked from the tiles and stored.

In order to minimise the overhead of data transfer between the host and device, input data and intermediate results stay resident where possible. For NX=16 and 800 elements, the initial \emph{u} data transfer is 6.4MB, with \emph{D} and \emph{$D^T$} both being 2KB. Results \emph{w} are also 6.4MB, however, the phase two transposition requires three 6.4MB transfers (\emph{u transposed}, \emph{ut transposed} and \emph{ut}) which adds to the host-device data movement considerably and both are stalled whilst this occurs. Elements are independent of each other and-so the processing of them is embarrassingly parallel. We exploit this property when packing into tiles, but this also enables us to decompose on an element by element basis across Tensix cores, with a subset of elements processed by each core. 

\section{Initial performance}
\label{sec:initial}
We undertook an initial performance comparison of our ported Nekbone AX kernel on a Wormhole n300 and compared it to a single core of a Xeon Platinum Cascade Lake (8260M). During the preliminary experiments, all code was compiled at optimisation level three, with the CPU code compiled using GCC 7.4, and for the Wormhole the host code was compiled using Clang 20 and the device code using TT-Metal v0.65. Results are averaged over five runs, and we ignore the first run which takes considerably longer on the Wormhole as the TT-Metal framework JIT compiles the code on first execution before launching. The Wormhole card is installed in an Intel Xeon Platinum Skylake (8170) system with PCIe Gen3. To obtain the power draw figures, the \emph{tt-smi} utility was used for Wormhole which returns the ASIC core power draw in Watts, and the Intel’s Running Average Power Limit (RAPL) was used for the CPU. It should be highlighted that whilst the Wormhole supports BF16 natively, the 8260M CPU does not and hence we compare against FP32 and FP64 on that architecture.

Table \ref{fig-initial-performance} reports performance for NX=16 and 800 elements. We use datatype BF16 on the Wormhole and FP32 on the CPU core with these being the fastest number representations on each technology. We report results on the Wormhole for 1 Tensix core, 64 Tensix cores (one chip on the n300) and 128 Tensix cores (two chips on the n300). It can be observed from Table \ref{fig-initial-performance} that irrespective of this, performance on the Wormhole is less than that of a single CPU core. Whilst the average power draw of the Wormhole is around four times less than the CPU core, the poor performance results in a power efficiency of up to around three times that of the CPU core. Furthermore, scaling the number of Tensix cores has a very limited performance improvement.

\begin{table}[htbp]
\centering
\begin{tabular}{ | c c c c | }
\hline
\textbf{Description} &  \makecell{\textbf{Performance} \\ \textbf{(GFLOPS)}} &  \makecell{\textbf{Power draw} \\ \textbf{(Watts)}} &  \makecell{\textbf{Power Efficiency} \\ \textbf{(GFLOPS/Watt)}} \\\hline
1 Tensix & 2.68 & 13.20 & 0.20 \\
64 Tensix & 3.50 & 15.70 & 0.22 \\
128 Tensix & 2.90 & 21.40 & 0.14 \\
1 core of CPU & 5.38 & 65.16 & 0.08 \\
\hline
\end{tabular}
\vspace{1mm}
\caption{Initial performance and power comparison between Wormhole (BF16) and CPU core (FP32) with NX=16 and 800 elements.}
\label{fig-initial-performance}
\end{table}

To understand the reason for the poor Wormhole performance reported in Table \ref{fig-initial-performance}, we scaled the number of elements across 128 Tensix cores, effectively varying the problem size. It was our initial hypothesis that there was not enough work for the Tenstorrent accelerator to perform, with kernels completing very quickly and performance being dominated by the overhead of data transfer between the host and device. Table \ref{fig-els-performance} reports a performance and power comparison for different numbers of elements on 128 Tensix cores at BF16. It can be observed that increasing the problem size does somewhat improve performance, however, this is not substantial and for the highest performance the Wormhole is still only just out-performing a single CPU core.

\begin{table}[htbp]
 \centering
\begin{tabular}{ | c c c c | }
\hline
\textbf{Elements} &  \makecell{\textbf{Performance} \\ \textbf{(GFLOPS)}} &  \makecell{\textbf{Power draw} \\ \textbf{(Watts)}} &  \makecell{\textbf{Power Efficiency} \\ \textbf{(GFLOPS/Watt)}} \\\hline
800 & 2.90 & 21.40 & 0.14 \\
5000 & 5.73 & 21.60 & 0.27 \\
10000 & 5.81 & 21.70 & 0.27 \\
100000 & 6.02 & 21.50 & 0.28 \\
\hline
\end{tabular}
\vspace{1mm}
\caption{Performance and power comparison of different problem sizes across 128 Tensix cores (BF16).}
\label{fig-els-performance}
\end{table}

We then measured the runtime for each of the three phases that were described in Section \ref{sec:porting} and this is reported in Table \ref{fig-phase-performance}, again for different number of elements which varies the problem size. It can be seen that the runtime is dominated by phase two, and indeed the runtime of phases one and three is fairly constant irrespective of the problem size. Clearly, the additional data transfer, data packing and unpacking, format conversion and transposition on the host is a significant bottleneck and a limitation of our initial design. Conversely, the design of phases one and three suits the pipelined streaming nature of the Tenstorrent architecture and scales well as the problem size is increased.

\begin{table}[htbp]
 \centering
\begin{tabular}{ | c c c c | }
\hline
\textbf{Elements} &  \makecell{\textbf{Phase one} \\ \textbf{(ms)}} &  \makecell{\textbf{Phase two} \\ \textbf{(ms)}} &  \makecell{\textbf{Phase three} \\ \textbf{(ms)}} \\\hline
800 & 18 & 177 & 23 \\
5000 & 19 & 648 & 22 \\
10000 & 19 & 1324 & 15 \\
100000 & 32 & 13058 & 23 \\
\hline
\end{tabular}
\vspace{1mm}
\caption{Runtime for each phase of the AX kernel on 128 Tensix cores (BF16) for different problem sizes.}
\label{fig-phase-performance}
\end{table}

\section{Optimisation of phase two}
\label{sec:optimise}
Based on the results described in Section \ref{sec:initial}, the design of phase two is suboptimal and results in significant overhead. In this baseline version, as part of phase two, the host transposes \emph{u} from (z,y,x) to (y,x,z), it then packs these transposed elements into 32x32 tiles and transfers them to the device. Once the phase has executed on the device the result field \emph{ut} is copied back to the host, unpacked back to a per-element basis, converted from BF16 to FP64, and transposed back to (z,y,x). At this point values are converted back to BF16, repacked into tiles and copied back to the device. Each of these stages allocates a fresh intermediate buffer and for NX=16 with 800 elements involves iterating over 3.3 million points.  

To optimise phase two, we explored two potential approaches; firstly, the fusion of these activities on the host, and secondly, moving both transpositions onto the device to entirely avoid involvement of the host here. The first approach, development of a fused variant, combined each pre-device and post-device execution stage in order to operate on elements in-place. Table \ref{fig-optimised-performance} reports the performance of this version, \emph{host fusion}, for different numbers of elements. It can be seen that, compared to the baseline code, this optimisation delivers around double the performance. The reason for this performance improvement is twofold. Firstly, the number of full passes over the data is reduced to one in each direction, cutting host-side memory traffic roughly in proportion. Because the working set, 6.4 MB of BF16 tiles for 800 elements, exceeds L2 cache, each eliminated pass avoids a round-trip to main memory and-so is potentially significant. Secondly, the reverse transposition path no longer converts the resulting BF16 numbers in \emph{ut} to FP64. In doing this the baseline version produces 26 MB of doubles for 800 elements, transposes them, and then converts back to BF16. By contrast, our fused kernel operates entirely in BF16 requiring two bytes per value instead of eight, quartering the memory footprint, which is crucial for caching, and removing two implicit type-conversion loops. Whilst the CPU operating on BF16 tends to be slower than FP64 \cite{ccgridbrown} because this number format does not tend to be supported in hardware, and-so arithmetic is undertaken in software, crucially here we are only undertaking data movement, and-so as there is no calculation the CPU's arithmetic support for the number format is not relevant. 

However, this fused approach still requires three data transfers between the host and device, and given that our host system only provides PCIe Gen3, this is especially expensive. When phase two is executed, the \emph{u} field is already present on the device, as it was transferred as part of phase one and stays resident as it will be used for phase three. Consequently, we developed a device-side transposition approach which transposes the resident \emph{u} field on the device, computes with this as part of phase two, and then performs a transposition on the result \emph{ut} back to (z,y,x) orientation which can then be directly consumed by phase three. This requires two additional phases, 2a and 2c, which are executed before and after the existing phase 2 matrix multiplication respectively. Phase 2a reads each destination tile's 16 source tiles from DRAM into L1 scratch, gathers one row per source tile to then assemble the transposed tile, and writes this back. Phase 2c performs the symmetric reverse gather. In addition to eliminating the three host-device data transfers, this version also avoids the host-side transposition loops which are single threaded and require a round trip to main memory. Crucially, this data reordering is able to leverage the FPU's transpose function and does not require any reordering by the data movement baby RISC-V core, which would be far slower. 

Table \ref{fig-optimised-performance} reports the performance of this optimisation, \emph{device transposition}, and it can be seen that for larger number of elements this very significantly improves performance compared to the baseline or fused approach. Effectively, the performance limitation has shifted from PCIe bandwidth to on-device DRAM to L1 NoC bandwidth, which is an order of magnitude greater than that of PCIe. However, the downside with this approach is that each destination tile requires reading NX source tiles to perform the transposition, which for a polynomial order of 16 is a sixteen times increase in reads compared to a stride-one copy. However, this is offset by the fact that on the Wormhole we are able to pipeline reads through the NoC and distribute them across the Tensix cores in parallel.

\begin{table}[htbp]
 \centering
\begin{tabular}{ | c c c c | }
\hline
\textbf{Elements} & \makecell{\textbf{Baseline} \\ \textbf{(GFLOPS)}} & \makecell{\textbf{Host fusion} \\ \textbf{(GFLOPS)}} &  \makecell{\textbf{Device transpose} \\ \textbf{(GFLOPS)}} \\\hline
800 & 2.90 & 5.54 & 5.59 \\
5000 & 5.73 & 11.09 & 45.91 \\
10000 & 5.81 & 11.96 &  91.81 \\
100000 & 6.02 & 13.69 & 242.97 \\
\hline
\end{tabular}
\vspace{1mm}
\caption{Performance comparison for baseline and two optimised versions for different problem sizes across 128 Tensix cores (BF16).}
\label{fig-optimised-performance}
\end{table}

Table \ref{fig-final-performance-800} reports the performance and power of our optimised Nekbone AX kernel for NX=16 and 800 elements on the Wormhole's 128 Tensix cores (BF16) against the Xeon Platinum CPU, an Nvidia V100 GPU and AMD U280 FPGA (all FP32) from \cite{brown2020exploring}. We leverage FP32 on the V100 GPU and U280 FPGA because, similarly to the Xeon Platinum CPU, neither supports BF16. It can be observed that the performance on the Wormhole is now comparable to that of a single CPU core and the power draw of the Wormhole is around three times less than that of the CPU core. Despite a better power efficiency than a single CPU core, for 800 elements the Wormhole is still not competitive against the 24-core Xeon CPU, GPU and the FPGA.

\begin{table}[htbp]
\centering
\begin{tabular}{ | c c c c | }
\hline
\textbf{Description} &  \makecell{\textbf{Performance} \\ \textbf{(GFLOPS)}} &  \makecell{\textbf{Power draw} \\ \textbf{(Watts)}} &  \makecell{\textbf{Power Efficiency} \\ \textbf{(GFLOPS/Watt)}} \\\hline
128 Tensix & 5.59 & 21.30 & 0.26 \\
1 core of CPU & 5.38 & 65.16 & 0.08 \\
24 cores of CPU & 65.74 & 176.65 & 0.37 \\
V100 GPU & 407.62 & 173.63 & 2.34 \\
U280 FPGA & 289.02 & 71.98 & 4.02 \\
\hline
\end{tabular}
\vspace{1mm}
\caption{Performance and power comparison between Wormhole (BF16) and the CPU, GPU and FPGA (all FP32) with NX=16 and 800 elements. CPU results compiled with GCC 7.4.}
\label{fig-final-performance-800}
\end{table}

Table \ref{fig-final-performance-800-10000-100000} reports the final performance and power of our optimised Nekbone AX kernel for NX=16 and 800, 10000 and 100000 elements on the Wormhole's 128 Tensix cores at BF16 against the 24-core Xeon Platinum CPU at FP32 and FP64. Here the CPU baseline was compiled with the newer GCC 10.2, rather than the GCC 7.4 used for the earlier results. There is a very significant performance difference on the Wormhole between 800 and 100000 elements, with the increased problem size ameliorating the overhead of host-device data communication for \emph{D}, \emph{$D^T$}, \emph{u} and \emph{w}, as well as suiting the pipelined nature of the architecture. 

\begin{table}[htbp]
\centering
\resizebox{0.95\columnwidth}{!}{%
\begin{tabular}{ | c c c c c c | }
\hline
\textbf{Description} & \textbf{Precision} & \textbf{Elements}
    & \makecell{\textbf{Performance}\\\textbf{(GFLOPS)}}
    & \makecell{\textbf{Power}\\\textbf{(Watts)}}
    & \makecell{\textbf{Efficiency}\\\textbf{(GFLOPS/W)}} \\\hline
\multirow{3}{*}{128-core Tensix} & \multirow{3}{*}{BF16} & 800      & 5.59            & 21.30  & 0.26           \\
                                 &                       & 10000  & 91.81           & 21.40  & 4.29           \\
                                 &                       & 100000 & \textbf{242.97} & \textbf{21.40}  & \textbf{11.35} \\\hline
\multirow{6}{*}{24-core Xeon CPU} & \multirow{3}{*}{FP32} & 800      & 159.36 & 114.70 & 1.39 \\
                                  &                       & 10000  & 161.04 & 137.25 & 1.17 \\
                                  &                       & 100000 & 161.93 & 141.34 & 1.15 \\\cline{2-6}
                                  & \multirow{3}{*}{FP64} & 800      & 106.22 & 131.79 & 0.81 \\
                                  &                       & 10000  & 110.69 & 143.49 & 0.77 \\
                                  &                       & 100000 & 110.90 & 144.07 & 0.77 \\
\hline
\end{tabular}}
\vspace{1mm}
\caption{Final performance and power comparison between Wormhole (BF16) and CPU (FP32 \& FP64) with NX=16 across 800, 10000, 100000 elements. CPU results compiled with GCC 10.2.}
\label{fig-final-performance-800-10000-100000}
\end{table}

It can be observed that, whilst for 800 elements the Wormhole is not competitive, for a larger problem size it outperforms the 24-core Xeon Platinum CPU (parallelised via OpenMP \cite{brown2020exploring}) performance by 1.5--2.2 times against FP32 and FP64 respectively. The Wormhole draws the lowest average power compared to the other configurations, around 7 times less than the CPU. Consequently, for 100000 elements this combination of high performance and low power draw results in significant power efficiency gains compared to other configurations.

\section{Conclusions and further work}
\label{sec:conc}

In this paper, we explored the porting of a spectral element method via the Nekbone proxy-app to the Tenstorrent RISC-V accelerator. Describing the mapping of the AX kernel to the Tensix architecture, our hypothesis was that the heavy use of matrix multiplications in the AX kernel would suit the matrix multiplication engine, FPU, in the Tensix co-processor. 

The greatest challenge in porting the code to this architecture has been the different directions involved in the matrix multiplications across \emph{x}, \emph{y}, and \emph{z} direction gradient calculations. This is handled trivially on the CPU by the reordering of loop iterations, although it can result in poor cache reuse behaviour. By contrast, matrix multiplication on the Tensix architecture does not provide the same flexibility and our initial design of undertaking transposition on the CPU for the \emph{z} direction resulted in a high performance bottleneck.

The most effective optimisation we found was to keep data on the Tenstorrent device and perform transpositions, using separate kernels, in-place. Effectively this trades PCIe bandwidth for NoC bandwidth, and even though the approach requires more data accesses per element this is offset by the NoC's high performance and dedicated hardware support for transposition in the FPU. Our performance results demonstrated that on the Tenstorrent architecture performance increases with the problem size. Given the pipelined nature of the Tensix architecture, where the FPU is designed for processing a large number of pipelined tiles, this is unsurprising but it does demonstrate that the architecture best suits large problems which can be pipelined in this manner. 

The reduced power draw of the Tenstorrent accelerator, compared to the other hardware explored in this paper, is impressive. This is especially the case when compared to the U280 FPGA, where FPGAs are well known for drawing significantly less power than CPUs and GPUs for HPC workloads \cite{10027489}. In the future, we also plan to explore leveraging the vector unit, SFPU, as an alternative to the FPU that was used exclusively in this work. That would enable FP32 on the Wormhole, and it would also be interesting to see how the performance of the SFPU compares to the FPU. Furthermore, we plan on exploring the performance of this kernel across multiple Wormhole cards and the next generation Blackhole.

\section*{Acknowledgement}
The CPU runs in this paper ran on NextGenIO which was funded from the EU Horizon 2020 research and innovation programme under grant agreement No 671591. This research was funded by the HPC-R project (EP/Z533701/1). For the purpose of open access, the author has applied a Creative Commons Attribution (CC BY) licence to any Author Accepted Manuscript version arising from this submission.

\bibliographystyle{splncs04}
\bibliography{references.bib}

\end{document}